\documentclass[12pt]{article} 
\usepackage{amsmath,amssymb}

\makeatletter
\@addtoreset{equation}{section}
\makeatother

\newcommand{\e }{\varepsilon }

\newcommand{\al }{\alpha }

\newcommand{\ket }{\rangle }
\newcommand{\bra }{\langle }

\newcommand{\Hw }{H_{\hbox{w}}} 
\newcommand{\im}{\hbox{Im}}
\newcommand{\re}{\hbox{Re}}

\newcommand{\no }{\nonumber }
\newcommand{\kob }{\overline{K^0}}
\newcommand{\ko }{K^0}
\newcommand{\dob }{\overline{D^0}}
\newcommand{\bob }{\overline{B^0}}
\newcommand{\bo }{B^0}

\newcommand{\nb }{\overline{n}}

\newcommand{\Ab }{\overline{A}}
\newcommand{\Gb }{\overline{G}}
\newcommand{\psib }{\overline{\psi}}

\begin{document}
\begin{flushright}
NUP-A-99-14\\July 1999
\end{flushright}
~\\ ~\\ 
\begin{center}
\Large{More on Parametrization Relevant to Describe Violation of \\
$CP$, $T$ and $CPT$ Symmetries in the $\ko$-$\kob$ System}
\end{center} 
~\\ ~\\ 
\begin{center}
Yutaka Kouchi, Yoshihiro Takeuchi\footnote{E-mail address: yytake@phys.cst.nihon-u.ac.jp} and S. Y. Tsai\footnote{E-mail address: tsai@phys.cst.nihon-u.ac.jp}
\end{center}

\begin{center}
{\it Atomic Energy Research Institute and Department of Physics \\ 
College of Science and Technology, Nihon University \\ 
Kanda-Surugadai, Chiyoda-ku, Tokyo 101-8308, Japan }
\end{center}
~\\ ~\\


\begin{abstract}

To study violation of $CP$, $T$ and/or $CPT$ symmetries in the $\ko$-$\kob$ systems, one has to parametrize the relevant mixing parameters and decay amplitudes in such a way that each parameter represents violation of these symmetries in a well-defined way. Parametrization is of course not unique and is always subject to phase ambiguities. We discuss these problems with freedom associated with rephasing of final (or intermediate) as well as initial states taken into account. We present a fully rephasing-invariant parametrization and a particular rephasing-dependent parametrization, and give a couple of comments related to these and other possible parametrizations.

\end{abstract}

\newpage
\section{Introduction}
~~~~The $\ko-\kob$ system has been extremely unique in testing and exploring a number of fundamental laws and phenomena of nature, in particular those related to conservation and violation of discrete space-time symmetries. It is by now well established that $CP$ symmetry is violated in the $\ko-\kob$ system and that this violation is very small, i.e. at the level of $10^{-3}$ in  amplitude.$^{1,2,3}$ It is expected that far more precise tests of this and related symmetries will be conducted at, say, $\phi$ factories.$^{4}$

To study possible violation of $CP$, $T$ and/or $CPT$ symmetries in the $\ko$-$\kob$ system, one has to parametrize the relevant mixing parameters and decay amplitudes in such a way that each parameter represents violation of these symmetries in a well-defined way. Parametrization is of course not unique and is always subject to phase ambiguities.

In a series of papers,$^{5,6,7,8,9}$ which are devoted to phenomenological studies of $CP$, $T$ and $CPT$ violations in the $\ko-\kob$ system, we have parametrized relevant mixing parameters and decay amplitudes in a way which is manifestly invariant with respect to rephasing of the $\ko$ and $\kob$ states. We have however adopted from the outset a specific phase convention as for final (or intermediate) states. As results, our arguments there appear not general enough. Although we have taken rephasing of final states into account in a subsequent report,$^{10}$ the arguments given contain something misleading. In this note, by taking into account freedom associated with rephasing of both initial and final states in a more careful and thorough way, we like to present a fully rephasing-invariant parametrization and a particular rephasing-dependent parametrization, and give a coule of comments related to these and other possible parametrizations.\footnote{A part of the contents of the present note was reported by one of the authors (S.Y.T) at the 8'th B Phisics International Workshop held at Kawatabi, Japan on October 29-31, 1998.$^{11}$}

\section{The $\ko$-$\kob$ mixing}
~~~~Let $|\ko\ket$ and $|\kob\ket$ be eigenstates of the strong interaction with strangeness $S = \pm 1$ related to each other by $CP$, $CPT$ and $T$ operations as$^{5,6,12}$
\begin{subequations}
\begin{eqnarray}
 CP|\ko\ket = e^{i\al_K}|\kob\ket~,&  &CPT|\ko\ket = e^{i\beta_B}|\kob\ket, \\
 CP|\kob\ket = e^{-i\al_K}|\ko\ket~,&  &CPT|\kob\ket = e^{i\beta_K}|\ko\ket, \\
 T|\ko\ket = e^{i(\beta_K-\al_K)}|\ko\ket~,&  &T|\kob\ket = e^{i(\beta_K+\al_K)}|\kob\ket. 
\end{eqnarray}
\end{subequations}
Note that, given Eq.(2.1a), Eqs. (2.1b) and (2.1c) follow from 
$(CP)^2 = (CPT)^2 = 1$, $CPT = (CP)T = T(CP)$ and antilinearity of 
$T$ and $CPT$ operations. $\al_K$ and $\beta_K$ are arbitrary real parameters to be referred to respectively as relative $CP$ and $CPT$ phases (between $|\ko\ket$ and $|\kob\ket$).

When the weak interaction $\Hw$ is switched on, $|\ko\ket$ and $|\kob\ket$ transit into 
other states, generically denoted as $|n\ket$, and get mixed. 
As solutions of the eigenvalue problem$^{13,14}$
\begin{equation}
\Lambda|~~ \ket = \lambda|~~ \ket,
\end{equation}
where $\Lambda = M -i\Gamma/2$ with $M$ ($\Gamma$) being the $2\times 2$ mass (width) matrix, one obtains two states $|K_S\ket$ and $|K_L\ket$ with definite mass $(m_{S,L})$ and width $(\gamma_{S,L})$:
\begin{subequations}
\begin{eqnarray}
|K_S\ket &=& (|p_S|^2+|q_S|^2)^{-1/2}(p_S|\ko\ket + q_S|\kob\ket), \\
|K_L\ket &=& (|p_L|^2+|q_L|^2)^{-1/2}(p_L|\ko\ket - q_L|\kob\ket),
\end{eqnarray}
\end{subequations}
with eigenvalues $\lambda_{S,L} = m_{S,L}-i\gamma_{S,L}/2$ given by
\begin{equation}
\lambda_{S,L} = \bra K_{S,L}|\Lambda|K_{S,L}\ket = 
(\bra \ko|\Lambda|\ko\ket + \bra \kob|\Lambda|\kob\ket)/2 \pm E,
\end{equation}
and ratios of the mixing parameters by
\begin{equation}
r_{S,L} \equiv q_{S,L}/p_{S,L} = \bra \kob|\Lambda|\ko\ket/ 
[E\pm(\bra \ko|\Lambda|\ko\ket - \bra \kob|\Lambda|\kob\ket)/2],
\end{equation}
where
$$
E = [\bra \ko|\Lambda|\kob\ket \bra \kob|\Lambda|\ko\ket 
+ (\bra \ko|\Lambda|\ko\ket - \bra \kob|\Lambda|\kob\ket)^2/4]^{1/2}.
$$
\section{Decay amplitudes}
~~~~$\ko$ and $\kob$ (or $K_S$ and $K_L$) have many decay channels. Denoting final  states generically as $|n\ket$, and sometimes as $|\nb\ket$, we shall consider decay amplitudes
\begin{subequations}
\begin{eqnarray}
\bra n|\Hw|\ko\ket = A_n = |A_n|e^{i\psi_n} ~, \\
\bra \nb|\Hw|\kob\ket = \Ab_{\nb} = |\Ab_{\nb}|e^{i\psib_{\nb}} ~.
\end{eqnarray}
\end{subequations}
$|n\ket$ and $|\nb\ket$ are, by definition, related to each other by $CP$, $CPT$ and $T$ operations as
\begin{subequations}
\begin{eqnarray}
 CP|n\ket = e^{i\al_n}|\nb\ket~,&  &CPT|n\ket = e^{i\beta_n}|\nb\ket~, \\
 CP|\nb\ket = e^{-i\al_n}|n\ket~,&  &CPT|\nb\ket = e^{i\beta_n}|n\ket~, \\
 T|n\ket = e^{i(\beta_n-\al_n)}|n\ket~,&  &T|\nb\ket = e^{i(\beta_n+\al_n)}|\nb\ket, 
\end{eqnarray}
\end{subequations}
where $\al_n$ and $\beta_n$ are again arbitrary real parameters.

As $|n\ket$, we shall concentrate on $CP$ eigenstates $|\pm\ket$ (e.g., $|2\pi\ket$ and $CP$-odd part of $|3\pi\ket$) and semi-leptonic states, $|\ell\ket \equiv |\pi^- \ell^+ \nu _{\ell}\ket$ and $|\overline{\ell}\ket \equiv |\pi^+\ell^-\overline{\nu_{\ell}}\ket$, and it is understood that
\begin{equation}
|\nb\ket = |n\ket ~,~~ \al_n = 0/\pi ~, ~~ \Ab_{\nb} = \Ab_n ~,~~\psib_{\nb} = \psib_n ,~~~~\hbox{for}~~|n\ket = |\pm\ket ~.
\end{equation}
For $|n\ket = |\ell\ket$ and $|\overline{\ell}\ket$, we shall further consider the amplitude ratio
\begin{equation}
r_n = \frac{\bra n|\Hw|\kob\ket}{\bra n|\Hw|\ko\ket}~.
\end{equation}

\section{Rephasing}
~~~~Let us now consider phase transformation or rephasing:
\begin{subequations}
\begin{eqnarray}
|\ko\ket \to |\ko\ket' = e^{-i\xi_K}|\ko\ket ~, & &|\kob\ket \to |\kob\ket' = e^{-i\xi_{\overline{K}}}|\kob\ket = e^{i\xi_K}|\kob\ket ~, \\
|n\ket  \to |n\ket' = e^{-i\xi_n}|n\ket ~, & &|\nb\ket \to |\nb\ket' = e^{-i\xi_{\nb}}|\nb\ket ~.
\end{eqnarray}
\end{subequations}
Note that we have made use of the relation $\xi_{\overline{K}}=-\xi_K$, in Eq.(4.1a), which comes from the fact that $|\ko\ket$ and $|\kob\ket$ are eigenstates of the strong interaction with strangeness $S = \pm 1$ and this interaction conserves $S$.$^{13,14}$ As regards Eq.(4.1b), it is understood that
\begin{equation}
\xi_{\nb} = \xi_n~, \qquad \hbox{for}~~|n\ket = |\pm\ket ~.
\end{equation}

The phases of the decay amplitudes, $\psi_n$ and $\psib_{\nb}$, as well as the relative $CP$ and $CPT$ phases, $\al_K$, $\beta_K$, $\al_n$ and $\beta_n$, defined in the previous sections, are all not invariant in general under the rephasing defined here. In fact, if one defines
\begin{subequations}
\begin{eqnarray}
CP|\ko\ket' = e^{i\al'_K}|\kob\ket' ~,&  &CPT|\ko\ket' = e^{i\beta'_K}|\kob\ket' ~, \\
CP|n\ket' = e^{i\al'_n}|\nb\ket' ~,&  &CPT|n\ket' = e^{i\beta'_n}|\nb\ket' ~, \\
'\bra n|\Hw|\ko\ket' = A'_n = |A_n|e^{i\psi'_n} ~,&  &'\bra \nb|\Hw|\kob\ket' =  \Ab'_{\nb} = |\Ab_{\nb}|e^{i\psib'_{\nb}} ~,
\end{eqnarray}
\end{subequations}
one finds
\begin{subequations}
\begin{eqnarray}
\al'_K = \al_K - 2\xi_K ~,&  &\beta'_K = \beta _K ~, \\ 
\al'_n = \al_n - \xi_n + \xi_{\nb} ~,&  &\beta'_n = \beta _n + \xi_n + \xi_{\nb} ~, \\
\psi'_n = \psi_n - \xi_K + \xi_n ~,&  &\overline{\psi}'_{\nb} = \overline{\psi}_{\nb} + \xi_K + \xi_{\nb} ~.
\end{eqnarray}
\end{subequations}
From these equations, one readily see that the particular combinations $\psi_n + (\beta_K - \al_K - \beta_n +\al_n)/2$ and $\psib_n + (\beta_K + \al_K - \beta_n - \al_n)/2$ (or $\psib_{\nb} -\psi_n + \al_K -\al_n$ and $\psib_{\nb} + \psi_n + \beta_K - \beta_n$), as well as $\beta_K$, are rephasing-invariant. One furthermore observes that it is possible to adjust $\xi_K$, $\xi_n - \xi_{\nb}$ and $\xi_n + \xi_{\nb}$ so as to have
\begin{subequations}
\begin{eqnarray}
\al_K' = 0~, &  & \\
\al_n' = 0~, &  &~~\hbox{for}~~ |n\ket = |\ell\ket~, \\
\psib'_{\nb} + \psi'_n = 0~~\hbox{or}~~ \beta_n' = \beta_K'~, &  &~~\hbox{for}~~ |n\ket = |\pm\ket~, ~~|\ell\ket~,
\end{eqnarray}
\end{subequations}
independent of one another.

The amplitude ratio, $r_n$, as well as the ratios of the mixing parameters, $r_{S,L}$, are independent of rephasing of the final (or intermediate) states $|n\ket$, Eq.(4.1b), and one may readily convince himself that $r_n e^{i\al_K}$, as well as $r_{S,L}e^{-i\al_K}$, are invariant under rephasing of the initial states $|\ko\ket$ and $|\kob\ket$, Eq.(4.1a).
\section{Conditions imposed by $CP$,~$T$ and $CPT$ symmetries}
~~~~From our definition of $CP$,~$T$ and $CPT$ transformations, Eqs.(2.1) and (3.2), we see that $CP$,~$T$ and $CPT$ symmetries impose on the decay amplitudes $A_n$ and $\Ab_{\nb}$ such conditions as
\begin{subequations}
\begin{eqnarray}
CP~~~&\to&~~ \Ab_{\nb} = A_n e^{-i(\al_K-\al_n)}~; \\
T~~~~&\to&~~ A_n^* = A_n e^{i(\beta_K-\al_K-\beta_n+\al_n)}~, \no \\
     &   &~~ \Ab_{\nb}^* = \Ab_{\nb}e^{i(\beta_K+\al_K-\beta_n-\al_n)}~; \\
CPT~~&\to&~~ \Ab_{\nb}^* = A_n e^{i(\beta_K-\beta_n)}~,
\end{eqnarray}
\end{subequations}
from which it follows that
\begin{subequations}
\begin{eqnarray}
CP~~~&\to&~~ |\Ab_{\nb}| = |A_n|~, \no \\
     &   &~~ \psib_{\nb} - \psi_n + \al_K -\al_n = 0~;\\
T~~~~&\to&~~ 2\psi_n + \beta_K - \al_K - \beta_n + \al_n = 0~, \no \\
     &   &~~ 2\psib_{\nb} + \beta_K + \al_K - \beta_n - \al_n = 0~;\\
CPT~~&\to&~~ |\Ab_{\nb}| = |A_n|~, \no \\
     &   &~~ \psib_{\nb} + \psi_n + \beta_K -\beta_n = 0~.
\end{eqnarray}
\end{subequations}
Note that all these are rephasing-invariant constraints. Similarly, one may readily see that $CP$, $T$ and $CPT$ symmetries impose on the amplitudes ratio, $r_n$, such conditions as
\begin{subequations}
\begin{eqnarray}
CP~~~&\to&~~ r_{\nb} = (1/r_n)e^{-2i\al_K}~; \\
T~~~~&\to&~~ r_n^* = r_n e^{2i\al_K}~; \\
CPT~~&\to&~~ r_{\nb}^* = (1/r_n)~.
\end{eqnarray}
\end{subequations}

As for the ratios of the mixing parameters, $r_{S,L}$, one may verify that $CP$,~$T$ and $CPT$ symmetries impose on the mass-width matrix $\Lambda$ such conditions as
\begin{subequations}
\begin{eqnarray}
CP~~~&\to&~~ \bra \ko|\Lambda|\ko\ket = \bra \kob|\Lambda|\kob\ket, \no \\
     &   &~~ \bra \ko|\Lambda|\kob\ket = \bra \kob|\Lambda|\ko\ket e^{-2i\al_K}~; \\
T~~~~&\to&~~ \bra \ko|\Lambda|\kob\ket = \bra \kob|\Lambda|\ko\ket e^{-2i\al_K}~; \\
CPT~~&\to&~~ \bra \ko|\Lambda|\ko\ket = \bra \kob|\Lambda|\kob\ket, 
\end{eqnarray}
\end{subequations}
and hence that
\begin{subequations}
\begin{eqnarray}
CP~~~&\to&~~ r_S = r_L = e^{i\al_K}~; \\
T~~~~&\to&~~ r_S r_L = e^{2i\al_K}~; \\
CPT~~&\to&~~ r_S = r_L~.
\end{eqnarray}
\end{subequations}
\section{A fully rephasing-invariant parametrization}
~~~~If one parametrize the amplitude $A_n$ and $\Ab_{\nb}$ as
\begin{subequations}
\begin{eqnarray}
A_n &=& G_n e^{-i(\beta_K-\al_K-\beta_n+\al_n)/2}~, \\ 
\Ab_{\nb} &=& \Gb_{\nb}e^{-i(\beta_K+\al_K-\beta_n-\al_n)/2}~,
\end{eqnarray}
\end{subequations}
it follows from our arguments given in Sec.4 and Sec.5 that both $G_n$ and $\Gb_{\nb}$ are rephasing-invariant and hence are complex in general, and that $CP$, $T$ and $CPT$ symmetries impose such conditions as
\begin{subequations}
\begin{eqnarray}
CP~~~&\to&~~ \Gb_{\nb} = G_n~; \\
T~~~~&\to&~~ G_n^* = G_n~, \qquad \Gb_{\nb}^* = \Gb_{\nb}~; \\
CPT~~&\to&~~ \Gb_{\nb}^* = G_n~.
\end{eqnarray}
\end{subequations}
If one parametrize $G_n$ and $\Gb_{\nb}$ further as
\begin{equation}
G_n = F_n(1+\e_n)~, \qquad \Gb_{\nb} = F_n(1-\e_n)~,
\end{equation}
one finds
\begin{subequations}
\begin{eqnarray}
CP~~~&\to&~~ \e_n = 0~; \\
T~~~~&\to&~~ \im(F_n) = 0~, \qquad \im(\e_n)= 0~; \\
CPT~~&\to&~~ \im(F_n) = 0~, \qquad \re(\e_n) =0~.
\end{eqnarray}
\end{subequations}
Similarly, if one parametrizes the amplitude ratios, $r_{\ell}$ and $r_{\overline{\ell}}$, as
\begin{subequations}
\begin{eqnarray}
r_{\ell} = x_{\ell} e^{-i\al_K}~,&  &r_{\overline{\ell}} = (1/x_{\overline{\ell}}^*) e^{-i\al_K}~, \\
x_{\ell} = x_{\ell}^{(+)} + x_{\ell}^{(-)}~,&  &x_{\overline{\ell}} = x_{\ell}^{(+)} - x_{\ell}^{(-)}~,
\end{eqnarray}
\end{subequations}
and the ratios of the mixing parameters, $r_{S,L}$, as
\begin{subequations}
\begin{eqnarray}
r_{S,L} = e^{i\al_K}\frac{1-\e_{S,L}}{1+\e_{S,L}}~, \\
\e_{S,L} = \e \pm \delta~,
\end{eqnarray}
\end{subequations}
one may verify that $x_{\ell}^{(+)}$, $x_{\ell}^{(-)}$, $\e$ and $\delta$ are all rephasing-invariant and that $CP$, $T$ and $CPT$ symmetries impose such conditions as
\begin{subequations}
\begin{eqnarray}
CP~~~&\to&~~ \re(x_{\ell}^{(-)}) = \im(x_{\ell}^{(+)}) = \e = \delta = 0~; \\
T~~~~&\to&~~ \im(x_{\ell}^{(+)}) = \im(x_{\ell}^{(-)}) = \e = 0~; \\
CPT~~&\to&~~ x_{\ell}^{(-)} = \delta = 0~.
\end{eqnarray}
\end{subequations}

Observed and/or expected smallness of violation of $CP$, $T$ and $CPT$ symmetries and of the $\Delta S = \Delta Q$ rule allows one to treat all the parameters, including $z_n \equiv \im(F_n)/\re(F_n)$ but excluding $\re(F_n)$, as small quantities.
\section{A partially rephasing-invariant parametrization}
~~~~It is legitimate to parametrize the amplitudes, $A_n$ and $\Ab_{\nb}$, in a rephasing-dependent way and at the same time adopt some phase convention.

We like to propose a particular rephasing-dependent parametrization, i.e., to parametrize $A_n$ and $\Ab_{\nb}$ as
\begin{subequations}
\begin{eqnarray}
A_n &=& \tilde{F}_n(1+\tilde{\e}_n)e^{i(\al_K-\al_n)/2}~, \\ 
\Ab_{\nb} &=& \tilde{F}_n(1-\tilde{\e}_n)e^{-i(\al_K-\al_n)/2}~,
\end{eqnarray}
\end{subequations}
but, as in the previous parametrization, leave $\al_K$ and $\al_{\ell}$ as well as $\beta_K$ and $\beta_n$ completely unspecified.

One may convince himself that $CP$, $T$ and $CPT$ symmetries impose on $\tilde{\e}_n$ such conditions as
\begin{subequations}
\begin{eqnarray}
CP~~~&\to&~~ \tilde{\e}_n = 0~; \\
T~~~~&\to&~~ \im(\tilde{\e}_n)= 0~; \\
CPT~~&\to&~~ \re(\tilde{\e}_n) =0~,
\end{eqnarray}
\end{subequations}
but do not impose any condition on $\tilde{F}_n$. One may however, by a choice of phase convention, set
\begin{equation}
\psib_{\nb} + \psi_n = 0~,
\end{equation}
which is equivalent to fix $\tilde{F}_n$ in such a way as
\begin{equation}
\im(\tilde{F}_n) = 0~.
\end{equation}
Here we have treated $\tilde{\e}_n$ as first-order small.
\section{Comments}
~~~~A couple of comments are in order.

(1) As mentioned in Sec.4, the chice (7.3) and the other two choices
\begin{subequations}
\begin{eqnarray}
\al_K = 0~, \\
\al_{\ell} = 0~,
\end{eqnarray}
\end{subequations}
are compatible with one another. Our parametrization (7.1) is general enough to accommodate one or both of the latter two choices.\footnote{We have adopted (7.3) and (8.1b) but left $\al_K$ unspecified in Ref.[11].}

(2) In our previous papers,$^{5,6,7,8,9}$ we have adopted the phase convention:
\begin{subequations}
\begin{eqnarray}
\al_{\ell} = 0~, \\
\beta_n = \beta_K.
\end{eqnarray}
\end{subequations}
Although we have preferred not to specify $\al_K$, we have noticed that it is legitimate to set $\al_K = 0$. The fully rephasing-invariant parametrization given in Sec.6 is general enough to accommodate all these phase conventions.

(3) The phase convention (7.3) and the phase convention (8.2b) are not compatible with each other. Applying these two phase conventions simultaneously to Eq.(6.1) and Eq.(6.3) would yield
\begin{equation}
\im(F_n) = 0~,
\end{equation}
a constraint which would follow only when $T$ and/or $CPT$ symmetries were exact (see Eq.(6.4)).

(4) Each of the two parametrizations given has its advantage and disadvantage. The parametrization given in Sec.6 has advantage that it is fully rephasing-invariant, but has disadvantage that some of the parameters involved may not be separately determinable. The parametrization given in Sec.7, in contrast, has disadvantage that it is not fully rephasing-invariant but rather only partially rephasing-invariant, but has advantage that the number of the parameters may be reduced by one for each mode by a choice of phase convention. 

(5) In connection with (3) and (4), we recall that we have previously encountered a similar situation.$^{5,6}$ For the case of $|\nb\ket = |n\ket = |2\pi\ket$, we find that our parameter $\im(\e_n)$ comes into play always in combination with $\im(\e)$, which renders these two parameters not separately determinable. In contrast, by parametrizing $r_{S,L}$ and $\Ab_{\nb}/A_n$ in a way which is not invariant under rephasing (4.1a), 
\begin{equation}
r_{S,L} = \frac{1-(\e'\pm\delta')}{1+(\e'\pm\delta')}~, \qquad \Ab_{\nb}/A_n = \frac{1-\e'_n}{1+\e'_n}~,
\end{equation}
one may set either $\im(\e') = 0$ or $\im(\e'_n) = 0$ by a choice of phase convention, rendering the other determinable. The phase convention $\im(\e'_n) = 0$, say, is not compatible with the phase convention $\al_K = 0$ and simultaneous adoption of these two phase conventions would yield $\im(\e_n) = 0$, a constraint which would follow only when $CP$ and/or $T$ symmetries were exact (see Eq.(6.4)).\footnote{Issues related to phase ambiguities are rather subtle and controversial. In fact, while some people$^{15,16,17,18}$ share basically the same viewpoints with us, some others possess different viewpoints.$^{19,20,21,22}$ Rephasing invariance has also been discussed from a somewhat different context by Kurimoto.$^{23}$}

In conclusion, we like to remark that our discussion on the $\ko-\kob$ system applies equally to the $D^0-\dob$ and $\bo-\bob$ systems, except that one needs to consider different kinds of final states for the latter systems.$^{10,24}$

\end{document}